# ENUMERATION OF *SALMONELLA* IN COMPOST MATERIAL BY A NON-CULTURE BASED METHOD

SUNAR, N. M.*+, STEWART, D.I *, STENTIFORD, E.I.*, AND FLETCHER, L.A.*

* *Pathogen Control Engineering (PaCE) Institute, School of Civil Engineering, University of Leeds, Leeds, LS2 9JT, United Kingdom.*
+ *Corresponding Author. Tel: +44 (0) 113 3432319. Email: shuhaila@uthm.edu.my*

SUMMARY: Accurate enumeration of *Salmonella* spp. is important for assessing whether this pathogen has survived composting. Recent literature has reported that enumeration of *Salmonella* spp. using standard microbiological methods has a numbers of disadvantages, particularly the time taken to obtain a result. This research is an attempt to develop a rapid, low-cost detection method that is quantitative, highly sensitive and target specific. This paper reports the development of a DNA fragment that can be used to quantify *Salmonella* spp. by competitive polymerase chain reaction (cPCR) targeted at the *invA* gene of *Salmonella* (PCR primers that target the *invA* gene are reported to have very high specificity for *Salmonella* strains). It is shown that cPCR, which could be completed in 5 hours, could quantify the number of copies of the *Salmonella invA* gene in a sample solution.

## 1. INTRODUCTION

Composting is an aerobic, biological process that uses naturally occurring microorganisms to convert organic waste into a humus-like product. This process is designed to both sanitise and stabilize the organic material (Imbeah, 1998). It is an environmentally sound process which is gaining worldwide popularity and is of considerable economic importance (Franke-Whittle et al., 2005). The pathogen content in compost is important because, if not properly treated the compost could be a potential source for pathogen dispersal into the environment. Research work with municipal wastes (Déportes et al., 1998, Hassen et al., 2001), sewage sludges (Dudley et al., 1980), and other organic sludges (Bustamante et al., 2008) has shown that they can contain a wide range of pathogens. *Salmonella* spp. and *E. coli* are widely used as pathogen indicators, and are included in many compost standards. Typical levels required of *Salmonella* spp. and *E. coli* in many European standards are absence in 25g and less than $10^3$ colony forming units (cfu).$g^{-1}$, respectively (Commission Decision 2001/688/EC; Commission Decision 2005/384/EC; and Briancesco, 2008). The UK composting "standard" PAS 100 (BSI, 2005) requires composts that are sold to be free of *Salmonella* spp. and to contain fewer than 1000 cfu of *E. coli* per gram of material.

The standard cultivation methods typically take 2-3 days to quantify the levels of indicators in compost (Sidhu et al., 2001, Déportes et al., 1998). These methods also have several disadvantages. In addition to being time-consuming, they are not always specific, omit organisms which are viable in their natural environment but can not be cultured (VBNC), and sometimes fail to detect *Salmonella* spp. when present (Novinscak et al., 2007, Lee et al., 2006, Kato and Miura, 2008).

Faced with these limitations, the intention of the work reported in this paper was to use a culture-independent, molecular-based method, to quantify the number of *Salmonella* spp. in a sample. The technique used was competitive polymerase chain reaction (cPCR) where the target DNA sequence is co-amplified with known amounts of a competitor DNA that shares most of its sequence with the target. As the target sequence and competitor fragment are amplified at that same rate, their relative abundance in the PCR product is a measure of their initial relative abundance. This paper reports the development of a suitable competitor DNA fragment for use in cPCR experiments targeted at a fragment of the *invA* gene of *Salmonella* spp. The competitor was constructed by deletion of 100bp from the target fragment without changing its end sequences, such that the competitor and target are amplified by the same PCR primers.

## 2. BACKGROUND

### 2.1 Conventional Procedure for Monitoring Composting

The conventional method that is used for the enumeration of *Salmonella* spp. is serial dilution followed by standard membrane filtration as recommended in the compost quality standard method PAS 100 (BSI, 2005) and the British Standard (BS EN ISO 6579:2002). A 25g sub-sample of compost is taken, placed into a sterile stomacher bag together with 225ml of sterile phosphate buffered saline solution and stomached for 60 seconds. The resulting suspension is then serially diluted and an aliquot of each dilution is filtered through a 0.45um cellulose nitrate filter. Each filter is then placed onto an absorbent pad onto which has been placed 1.5ml of resuscitation medium (tetrathionate broth). The pads are then incubated for 18-24 hours at 37°C. Each filter is then transferred to a sterile Petri dish containing Rambach agar and incubated for a further 24 hours at 37°C after which the resulting colonies are counted.

### 2.2 The Polymerase Chain Reaction

Polymerase chain reaction (PCR) is a technique that amplifies (i.e. replicates) a single fragment, or a few copies of a fragment of DNA (deoxyribonucleic acid) to produce millions of copies of that particular DNA sequence. DNA is a double stranded molecule held together by hydrogen bonds between the complementary bases on the two strands (hydrogen bonds form between bases A&T and G &C). During PCR the target DNA is subjected to thermal cycling in an aqueous solution containing deoxynucleotide triphosphates (dNTPs; bases bound to deoxyribose), appropriate PCR primers (small starting fragments of DNA), and a thermo-stable DNA polymerase enzyme. Heating of the DNA separates the strands, cooling to an appropriate annealing temperature allows primers to bond to each DNA strand by formation of hydrogen bonds, incubation at an intermediate extension temperature allows synthesis of new DNA strands by the polymerase enzyme from the appropriate dNTPs. In principle each thermal cycle will double the number of copies of the target DNA fragment in the reaction.

In practice conventional PCR can not be used to quantify the amount of DNA in biological samples as the real amplification rate influenced by hard to control factors such as the relative abundance of starting material and the concentration of each dNTP remaining in solution, and a linear relationship the amount of template and PCR product is maintained only during the initial cycles of PCR amplification (Zentilin and Giacca, 2007).

Competitive PCR is a powerful technique for quantifying DNA that overcomes the limitations of conventional PCR by co-amplifying the target DNA fragment with a known amount of a competitor DNA fragment. If the competitor fragment is selected such that it shares most of its sequence with the target, particularly the primer annealing sites, then the two templates will be subject to the same predictable and unpredictable variables that affect amplification rate (Zentilin and Giacca, 2007). Thus the ratio of the target and competitor after PCR reflect the relative initial amounts of the two templates. Often the competitor is selected to be slightly shorter than the target so that the two fragments can be readily distinguished after PCR (e.g. by agarose gel electrophoresis). Quantification is more straight-forward if the concentration of target and competitor are similar, thus a cPCR experiment usually involves a series of reactions containing the same amount of target, and a serial dilution of a known amount of competitor.

## 3. METHODS

### 3.1 Selection of primers

PCR primers were selected that target a 285 bp segment of *invA* gene of *Salmonella* (the *invA* gene of *Salmonella* is a component of the cell invasion apparatus; Galan et al., 1992). These primers were *invA*139 (5' gtg aaa tta tcg cca cgt tcg ggc aa 3') and *invA*141 (5' tca tcg cac cgt caa agg aac c 3'), which target locations 287-312 and 571-550 within the *invA* gene, respectively (Rahn et al., 1992). These primers have been shown to have excellent specificity for *Salmonella*, detecting 99.4% *Salmonella* strains without false positives when tested against 630 *Salmonella* strains and 142 non-*Salmonella* strains (Rahn et al., 1992).

A PCR with these primers was conducted with DNA from *Salmonella enteritidis* (ATCC13076). A suspension of *Salmonella* cells in sterile distilled water were lysed by heating to 99°C for 5 minutes. Cell debris was removed by centrifugation. A PCR reaction was set-up containing 2.5 µl of DNA solution, 5 units of GoTaq DNA polymerase (Promega Corp., USA), 1× GoTaq PCR reaction buffer (containing 1.5mM $MgCl_2$), 0.2mM PCR nucleotide mix (Promega Corp., USA), and 0.6 µM DNA primers in a final volume of 50 µl. This reaction mixture (and a sterile control) was incubated at 95°C for 2 min, and then cycled 30 times through three steps: denaturing (95°C, 30s), annealing (50°C, 30s), primer extension (72°C, 45s). This was followed by a final extension step at 72°C for 7min. The PCR product was purified using agarose gel electrophoresis and a QIAquick Gel Extraction Kit (QIAGEN Ltd, UK). The PCR product, which was just under 300bp long, was sent for direct DNA sequencing (ABI 3100*xl* Capilliary Sequencer) using both the *invA*139 and *invA*141 primers.

### 3.2 Construction of the competitor fragment

A competitor DNA fragment with a small deletion was constructed from the DNA fragment produced by the primers *invA*139 and *invA*141 by following the outline protocol described by Zentilin and Giacca (2007). This procedure requires a third "internal" primer that consists of a ~20bp sequence that will anneal to a site within the DNA template attached to the 3′ end of one of the primers that produced that template. Design of this 20bp sequence was undertaken from the target sequence shown in Figure 1a using Primer3 (Rozen and Skaletsky, 2000) to produce a competitor fragment about ~100bp shorter than the target for easy identification by agarose gel electrophoresis. Two suitable 20bp sequences were identified (5′ ctg ttt acc ggg cat acc at 3′ and 5′ ggg cat acc atc cag aga aa 3′) and two primers (Sal-I and Sal-I-V1, respectively) were made by attaching these sequences to the 3′ end of *invA*141 (these primers were manufactured by Eurofins MWG Operon, Germany).

Competitor fragments were made by a PCR reaction using the *invA*139 and either the Sal-I or Sal-I-V1 primers. Each PCR contained 5 µl of DNA solution, 5 units of GoTaq DNA polymerase

(Promega Corp., USA), 1× GoTaq PCR reaction buffer (containing 1.5mM $MgCl_2$), 0.2mM PCR nucleotide mix (Promega Corp., USA), and 0.6 µM DNA primers in a final volume of 50 µl. The reaction mixtures (and a sterile control) were incubated at 95°C for 2 min, and then cycled 30 times through three steps: denaturing (95°C, 30 s), annealing (53°C, 30 s), primer extension (72°C, 45 s). This was followed by a final extension step at 72°C for 7min. The PCR products were visualised by agarose gel electrophoresis, and the gel bands representing the desired products were excised and purified using QIAquick Gel Extraction Kit (QIAGEN Ltd, UK).

The PCR products were ligated into a standard cloning vector (p-GEM-T Easy supplied by Promega), transformed into *E. Coli* cells (XL1-Blue supercompetent cells from Stratagene), and colonies were grown on LB- agar plates containing ampicillin (100 µg.ml$^{-1}$) surface dressed with IPTG and X-gal (as per the Stratagene protocol) for blue-white colour screening. Colonies containing the insert were re-streaked on LB-ampicillin agar plates, and single colonies from these plates were incubated overnight in liquid LB-ampicillin. Plasmid DNA was extracted using a QIAprep Spin minprep kit (QIAGEN Ltd, UK) and sent for automated DNA sequencing (ABI 3100*xl* Capilliary Sequencer) using the T7P primer.

After confirmation that the plasmid contained the correct insert, cells derived from the same original single colony were incubated overnight in liquid LB-ampicillin so that larger amounts of plasmid DNA could be extracted. The concentration of the plasmid in solution was measured using a NanoDrop ND-1000 UV-spectrophotometer (Thermo-Fisher Scientific Inc., USA). This concentration was converted to an approximate number of plasmid copies by the assumption 1µg of 1,000bp DNA contains $9.1 \times 10^{11}$ molecules (NEB, 2009) after correcting for the length of the plasmid formed by pGEM-T Easy and the insert.

### 3.3 cPCR reaction conditions

A cell culture from *Salmonella enteritidis* (ATCC13076) was lysed (99°C for 5 min), centrifuged to remove the cell debris, and the supernatant diluted by a factor of 10 was used as the target DNA solution. Each experiment consisted of a number of PCR reactions containing an equal amount of target DNA and one member of a dilution series of the competitor DNA (0, 20, 200, 2,000, 20,000 copies/µL). Each PCR reaction mixture contained 10 µL of target DNA solution, 2µL of competitor DNA solution, 5 units of GoTaq DNA polymerase (Promega Corp., USA), 1× PCR reaction buffer (containing 1.5mM $MgCl_2$), 0.2mM PCR nucleotide mix (Promega Corp., USA), and 0.6 µM DNA primers (*invA*139 and *invA*141) in a final volume of 50 µl. The reaction mixtures were incubated at 94°C for 4 min, and then cycled 40 times through three steps: denaturing (94°C, 30 s), annealing (53°C, 30 s), primer extension (72°C, 60 s). This was followed by a final extension step at 72°C for 7min. The PCR products were imaged by electrophoresis of 10 µl samples in a 1.5% agarose TBE gel with ethidium bromide straining.

### 3.4 DNA sequence analysis

Sequences analysis was undertaken in May 2009 using the NCBI-BLAST2 program and the EMBL release nucleotide database (Wéry et al., 2008). Default settings were used for the BLAST parameters (match/mismatch scores 2, -3, open gap penalty 5, gap extension penalty 2).

**(a) Fragment of the *invA* gene of *Salmonella enteritidis***
<u>GTGAAATTATCGCCACGTTCGGGCAA</u>TTCGTTATTGGCGATAGCCTGGCGGTGGGTTTTGTTGTCTTCTCTATTG
TCACCGTGGTCCAGTTTATCGTTATTACCAAAGGTTCAGAACGCGTCGCGGAAGTCGCGGCCCGATT<span style="border-bottom:1px dotted">TTCTCTGG</span>
**ATGGTATGCCCGGTAAACAG**ATGAGTATTGATGCCGATTTGAAGGCCGGTATTATTGATGCGGATGCTGCGCGCG
AACGGCGAAGCGTACTGGAAAGGGAAAGCCAGCTTTAC<u>GGTTCCTTTGACGGTGCGATGA</u>
Length 285bp

**(b) Deletion fragment made using the Sal-I primer**
<u>GTGAAATTATCGCCACGTTCGGGCAA</u>TTCGTTATTGGCGATAGCCTGGCGGTGGGTTTTGTTGTCTTCTCTATTG
TCACCGTGGTCCAGTTTATCGTTATTACCAAAGGTTCAGAACGCGTCGCGGAAGTCGCGGCCCGATTTTCTCTGG
**ATGGTATGCCCGGTAAACAG**GGTTCCTTTGACGGTGCGATGA
Length 192bp

**(c) Deletion fragment made using the Sal-I-V1 primer**
<u>GTGAAATTATCGCCACGTTCGGGCAA</u>TTCGTTATTGGCGATAGCCTGGCGGTGGGTTTTGTTGTCTTCTCTATTG
TCACCGTGGTCCAGTTTATCGTTATTACCAAAGGTTCAGAACGCGTCGCGGAAGTCGCGGCCCGATT<span style="border-bottom:1px dotted">TTCTCTGG</span>
ATGGTATGCCCGGTTCCTTTGACGGTGCGATGA
Length 183bp

**Key**
<u>*invA*139 primer</u>
<u>*invA*141 primer</u>
**sequence used to design Sal-I primer**
<span style="border-bottom:1px dotted">sequence used to design Sal-I primer</span>

**Figure 1:** (a) Fragment of the *invA* gene of *Salmonella* produced by the *invA*139 and *invA*141 primers showing the the primer binding locations exploited to produce the deletion fragments using (b) the Sal-I and (c) Sal-I-V1 internal primers.

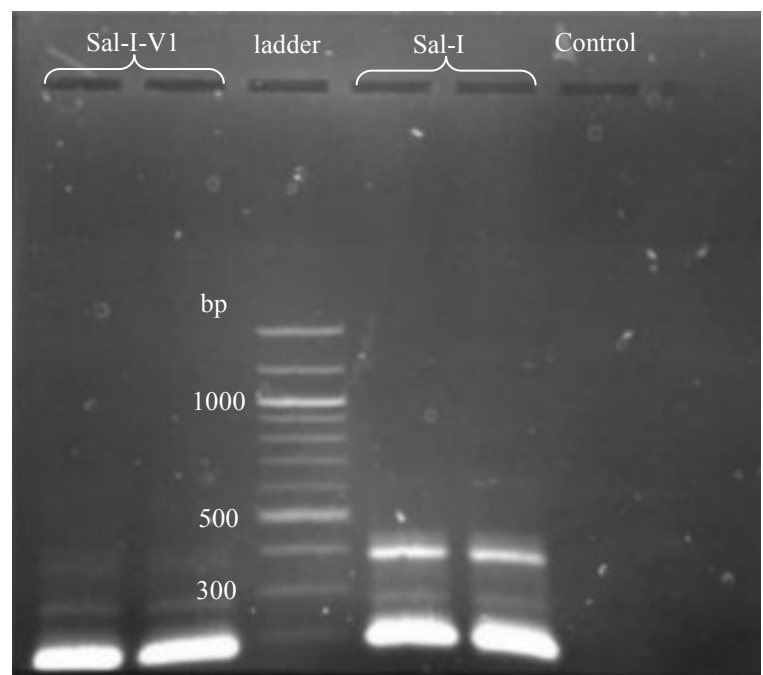

**Figure 2:** Gel image showing the competitor fragments (at just below 200bp) and other PCR products formed the Sal-I or Sal-I-V1 primers.

# 4. RESULTS

## 4.1 Production of a Competitor Fragment

A PCR with DNA from *Salmonella enteritidis* and the *invA*139 and *invA*141 primers yielded a 285bp fragment (see Figure 1a). This fragment was a perfect match to 13 *Salmonella* sequences contained within the EMBL Release database. Regions of the sequence corresponding to the *invA*139 and *invA*141 primers are doubly and singly underlined, respectively, in figure 1a. The DNA sequence corresponding to the 20bp sequence at the 3′ end of the Sal-I and Sal-I-V1 primers are shown in bold and with a dotted underline, respectively.

Further PCRs using the *invA*139 and either the Sal-I or Sal-I-V1 primers yielded a product that was around 200bp long in both cases (the desired product). Cloning and sequencing of these PCR products revealed that Sal-I yielded a 192 bp competitor fragment (Figure 1b) while Sal-I-V1 yielded a 183 bp competitor fragment (Figure 1c). As would be expected, both of these deletion fragments had end sequences that are compatible with the primers *invA*139 and *invA*141 (the primer sequences are highlighted on Figures 1b &c). Careful review of the gel image (Figure 2) also reveals gel bands with lengths of about 300 and 400 bp that are not the desired products. Such undesired bands are often produced by PCR when large amounts of DNA template are added to the reaction (as was the case here). The bands around 300 bp may represent unreacted template, which was 285bp long, but the bands around 400bp must result from dimer formation as no DNA of that length was added to the reactions (the control reaction indicates there was no contamination). The competitor fragment produced using the Sal-I-V1 primer was selected for use in the cPCR experiments because of the significantly higher propensity of the fragment produced using the Sal-I primer to form nonspecific PCR products.

## 4.1 cPCR

The outcome of the cPCR experiment can be seen in the gel image shown in Figure 3. All the reactions except for the control show very bright bands at just below 300 bp. These are assumed to be 285bp PCR product from the target *Salmonella* DNA. Samples SC400, SC4k and SC40k also exhibit a band at just below 200bp that increases with brightness with increasing concentration of the competitor fragment. These bands are assumed to be the PCR product from the 183bp competitor fragment. The equal brightness of the two main bands from sample SC40k indicates that once allowance is made for initial dilution 1µL of the *Salmonella* culture used to provide target DNA contained approximately 40,000 copies of the *invA* gene. Further work is ongoing to establish the relationship between the number of gene copies and the number of live cells in the original culture.

# 5. DISCUSSION

This pilot study has clearly demonstrated that cPCR has the potential to enumerate *Salmonella* species in aqueous solution. The 40-cycle programme used for cPCR takes approximately 2.5 hrs to run. The different bands on an agarose gel produced by the target and competitor DNA are easily distinguishable on a normal size gel after 1.5 hrs, thus the entire protocol can be undertaken in less than 5 hrs. Other reactions with very low target DNA copy numbers (not reported here) have shown that 40 copies of the competitor are readily detected by PCR as a bright band an agarose gel, and significantly fewer that 40 copies of the target DNA (i.e. a target concentration that produced no band when 40 copies of the competitor fragment were present) can be detected in the absence of the competitor fragment. Thus, provided a cPCR dilution series is always one PCR reaction with no competitor fragment, it is possible to detect extremely low copy numbers (possibly one or two copies) of the target.

The PCR programme used in this study for cPCR has been deliberately selected to maximise the probability of obtaining a product from very low copy numbers. Features include a relatively low annealing temperature of 53°C, and a large number of thermal cycles. The former will maximise the chances of the primers annealing to the DNA template, but may allow non-specific annealing (annealing at a site where there is some degree of mismatch). The outcome of such non-specific annealing would be non-specific PCR products which are usually easy to distinguish from the desired product due to their difference in length (it is extremely improbable that non-specific annealing of primer to either an incorrect site on *Salmonella* template or to non-*Salmonella* DNA will produce a product of exactly 183bp). Experiments to date with *Salmonella* as the only template do not indicate any issues with non-specific amplification, but further work with environmental samples are required to completely eliminate this issue. A PCR programme involving 40 thermal cycles is sufficient to reach the plateau phase of amplification for most systems, and thus the products will be readily detectable with rapid, low-cost, easy to use techniques like gel electrophoresis.

Current regulations require compost to be free from *Salmonella*, and thus it is debatable whether the expense of cPCR over conventional PCR will be justified outside of the research laboratory. However development of cPCR to enumerate *E.coli* in compost is being undertaken in parallel to the work reported here. As PCR detects the presence of pathogen specific DNA, rather than cell growth it will be able to identify the presence of VBNC cells that are not enumerated by traditional culturing methods (Higgins et al., 2007), there would appear to be a role for conventional PCR in compost testing.

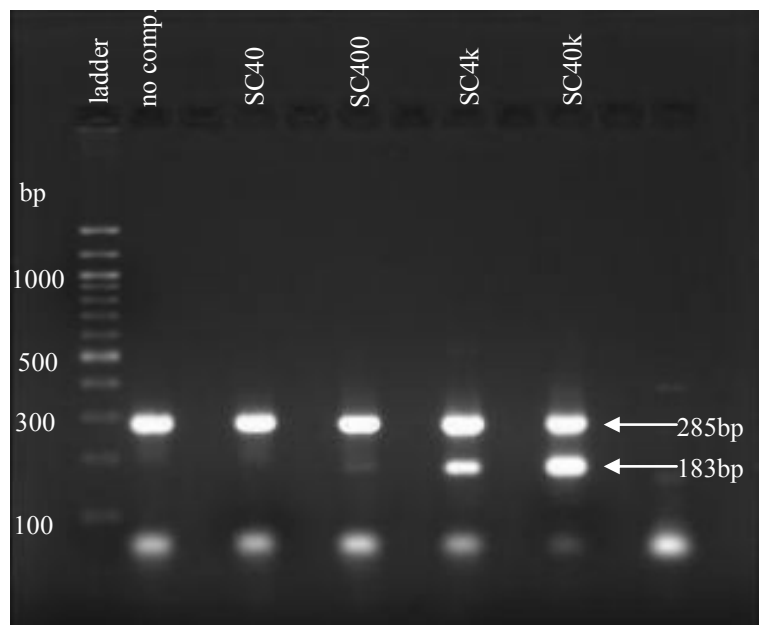

**Figure 3:** Gel image from a cPCR on a dilute Samonella culture. Reading from left to right the reactions contained zero, 40, 400, 4,000, and 40,000 copies of the 183bp competitor fragment, respectively.

The cPCR procedure detailed above does not differentiate between viable and dead cells, which means it could potentially overestimate the number of viable cells in a system. This would be a serious limitation for a test applied to sanitising process, such as composting, as dead cells may remain after the waste is safe. The next stage of this project is to incorporate an ethidium monoazide (EMA) pretreatment step into the cPCR protocol to provide viable/dead cell differentiation (Rudi et al. 2005). EMA, which is added to the system before cell lysis, can only penetrate cells with compromised membrane/cell wall systems where it will covalently bind to DNA after photoactivation. DNA covalently bound to EMA cannot subsequently be amplified by PCR.

## 6. CONCLUSION AND RECOMMENDATION

- Primers targeting the *invA* gene are recommended for detecting *Salmonella* spp. by PCR due to their high specificity.
- The deletion method of producing pathogen specific competitor fragments for use in cPCR has proved to be straight-forward, and has successfully been used to develop a competitor fragment for quantifying *Salmonella* spp.
- The cPCR technique can be used to enumerate *Salmonella* spp. in aqueous samples and should therefore be suitable for assessing pathogen removal in the composting process.
- Further work is required to fully validate the use of cPCR to enumerate pathogen in environmental samples, and to further develop the technique to differentiate between viable and dead cells.

ACKNOWLEDGMENT

This study was funded by Ministry of Higher Education of Malaysia and University of Tun Hussien Onn Malaysia. The authors wish to thank the Public Health Laboratory staff, School of Civil Engineering, University of Leeds for their valuable support and excellent laboratory facilities.

REFERENCES

Bej, A.K., Dicesare, J.L., Haff L., & Atl, R.M. (1991). Detection of Escherichia coli and Shigella spp. in Water by Using the Polymerase Chain Reaction and Gene Probes for uid. Appl. Environ. Microbiol., **57** 1013-1017.

Briancesco, R. Coccia, A.M., Chiaretti, G., Libera, S.D., Semproni, M., & Bonadonna, L., (2008). Assessment of microbiological and parasitological quality of composted wastes: health implications and hygienic measures. Waste Management & Research, **26**, 196-202

BS EN ISO 6579:2002. Microbiology of food and animal feeding stuffs. Horizontal method for the detection of *Salmonella* spp. British Standard Institute, London

BSI (2005) Specification for composted materials, Publicly Available Specification, PAS 100:2005, British Standard Institute, London.

Bustamante, M.A., Moral, R., Paredes, C., Vargas-Garcia, M.C., Suarez-Estrella, F. & Moreno, J. (2008). Evolution of the pathogen content during co-composting of winery and distillery wastes. Bioresource Technology, In Press, Corrected Proof.

Chen, Y-C, Higgins, M.J., Maas, N.A. & Murthy, S.N. (2006). DNA extraction and Escherichia

coli quantification of anaerobically digested biosolids using the competitive touchdown PCR method. Water Research, **40**, 3037-3044.

Dudley, D.J., Guentzel, M. N., Ibarra, M.J., Moore, B.E. AND Sagik B.P. (1980). Enumeration of potentially pathogenic bacteria from sewage sludges. Appl Environ Microbiol, **39**, 118-126.

Déportes, I., Zmirou B-G, & Bouvier, M-C (1998). Microbial disinfection capacity of municipal solid waste (MSW) composting. Journal of Applied Microbiology, **85**, 238-246.

Franke-Whittle, I.H., Klammer, S.H. & Insam, H. (2005). Design and application of an oligonucleotide microarray for the investigation of compost microbial communities. Journal of Microbiological Methods, **62**, 37-56.

Galan, J. E., Ginocchio, C. & Costeas, P. (1992) Molecular and functional characterization of the *Salmonella* invasion gene *invA*: homology of *invA* to members of a new protein family. J. Bacteriol., **174**, 4338-4349.

Grant, M.A., Weagant, S.D., & Feng, P. (2001). Glutamate Decarboxylase Genes as a Prescreening Marker for Detection of Pathogenic Escherichia coli Groups. Appl. Environ. Microbiol., **67**, 3110–3114.

Hassen, A., Belguith, K., Jedidi, N., Cherif, A., Cherif, M. & Boudabous, A. (2001). Microbial characterization during composting of municipal solid waste. Bioresource Technology, **80**, 217-225.

Higgins, M.J., Chen, Y-C, Murthy, S.N., Hendrickson, D., Farrel, J. & Schafer, P. (2007). Reactivation and growth of non-culturable indicator bacteria in anaerobically digested biosolids after centrifuge dewatering. Water Research, **41**, 665-673.

Imbeah, M. (1998) Composting piggery waste: A review. Bioresource Technology Volume **63**, 197-203.

Jin, C-F, Mata, M., & Fink, D.J. (1994). Rapid construction of deleted DNA fragments for use as internal standards in competitive PCR. Cold Spring Harbor Laboratory Press, www.genome.org downloaded on May 23, 2008.

Kato, K., & Miura, N., (2008). Effect of matured compost as a bulking and inoculating agent on the microbial community and maturity of cattle manure compost. Bioresource Technology **99**, 3372–3380.

Lee, D-Y, Shannon, K., Beaudette, L.A. (2006). Detection of bacterial pathogens in municipal wastewater using an oligonucleotide microarray and real-time quantitative PCR. Journal of Microbiological Methods, **65**, 453-467.

Mcdaniels, A.E., Rice, E.W, Reyes, A.L., Johnson, C.H., Haugland, R.A., Stelma, G.N. (1996). Confirmational identification of Escherichia coli, a comparison of genotypic and phenotypic assays for glutamate decarboxylase and β-D-glucuronidase. Appl. Environ. Microbiol. **62**, 3350-3354.

NEB (2009). New England Biolabs Catalogue and Technical Reference 2009-10. New England Biolabs, Inc.

Novinscak, A. Surette, C., & Filion, M., (2007). Quantification of *Salmonella* spp. in composted biosolids using a TaqMan qPCR assay. Journal of Microbiological Methods **70**, 119–126

Rahn, K., Degrandis, S.A., Clarke, R. C., Mcewen, S.A., Galan, J.E., Ginocchio, C., Curtiss, R. & Gyles, C.L. (1992). Amplification Of An *inva* Gene Sequence Of *Salmonella*-Typhimurium By Polymerase Chain-Reaction As A Specific Method Of Detection Of *Salmonella*. Molecular and Cellular Probes, **6**, 271-279.

Rozen, S. and Skaletsky, H.J. (2000). Primer3 on the WWW for general users and for biologist programmers. In: Krawetz S, Misener S (eds) Bioinformatics Methods and Protocols: Methods in Molecular Biology. Humana Press, Totowa, NJ, pp 365-386

Rudi, K., Moen, B., Dromtorp, S.M., & Holck, A.L. (2005). "Use of ethidium monoazide and PCR in combination for quantification of viable and dead cells in complex samples." Appl. Environ. Microbiol. **71**, 1018-1024.

Sabat, G., Rose, P., Hickey, W.J., & Harkin, J.M. (2000). Selective and sensitive method for

PCR amplification of Escherichia coli 16S rRNA genes in soil. Appl. Environ. Microbiol., **66**, 844-849

Shannon, K.E., Lee, D-Y, Trevors, J.T. & Beaudette, L.A. (2007). Application of real-time quantitative PCR for the detection of selected bacterial pathogens during municipal wastewater treatment. Science of the Total Environment, **382**, 121-129.

Sidhu, J., Gibbs, R.A., Ho, G.E. & Unkovich, I. (2001). The role of indigenous microorganisms in suppression of *Salmonella* regrowth in composted biosolids. Water Research, **35**, 913-920.

Tsai, Y., Palmer, C.J., & Sangermano, L.R., (1993). Detection of Escherichia coli in sewage and sludge by polymerase chain reaction. Appl. Environ. Microbiol., **59**, 353-357.

Watterworth, L., Topp, E., Schraft, H. & Leung, K.T. (2005). Multiplex PCR-DNA probe assay for the detection of pathogenic Escherichia coli. Journal of Microbiological Methods, **60**, 93-105.

Wéry, N., Lhoutellier, C., Ducray, F., Delgenès, J-P, & Godon, J-J, (2008). Behaviour of pathogenic and indicator bacteria during urban wastewater treatment and sludge composting, as revealed by quantitative PCR. Water Research, **42**, 53-62.

Zentilin, L. & Giacca, M. (2007). Competitive PCR for precise nucleic acid quantification. Nat. Protocols **2**(9): 2092.